\documentclass[a4paper,reqno]{amsart}
\usepackage{amssymb}
\usepackage{latexsym}
\usepackage{amsmath}
\usepackage{euscript}
\usepackage{graphics,color}
\usepackage[all]{xy}
\usepackage[margin=3cm]{geometry}
\usepackage{MnSymbol} 

\newcommand{\be}{\begin{equation}}
\newcommand{\ee}{\end{equation}}
\newcommand{\ba}{\begin{eqnarray}}
\newcommand{\ea}{\end{eqnarray}}
\newcommand{\baa}{\begin{eqnarray*}}
\newcommand{\eaa}{\end{eqnarray*}}
\newcommand{\bb}{\begin{thebibliography}}
\newcommand{\eb}{\end{thebibliography}}

\newcommand{\bi}[1]{\bibitem{#1}}
\newcommand{\lab}[1]{\label{#1}}
\newcommand{\re}[1]{(\ref{#1})}

\newcounter{my}
\newcommand{\he}%
   {\stepcounter{equation}\setcounter{my}%
   {\value{equation}}\setcounter{equation}0%
   }%
\newcommand{\she}%
   {\setcounter{equation}{\value{my}}%
    }%

\renewcommand\t{\tilde}

\newtheorem{pr}{Proposition}

\theoremstyle{definition}

\numberwithin{equation}{section}

\begin{document}

\title{The rational Heun operator and Wilson biorthogonal functions}

\author{Satoshi Tsujimoto}
\author{Luc Vinet}
\author{Alexei Zhedanov}

\address{Department of Applied Mathematics and Physics, Graduate School of Informatics, Kyoto University,
Yoshida-Honmachi, Kyoto, Japan 606-8501}

\address{Centre de recherches \\ math\'ematiques,
Universit\'e de Montr\'eal, P.O. Box 6128, Centre-ville Station,
Montr\'eal (Qu\'ebec), H3C 3J7}

\address{School of Mathematics, Renmin University of China, Beijing 100872,CHINA}

\begin{abstract}
We consider the rational Heun operator defined as the most general second-order $q$-difference operator which sends any rational function of type $[(n-1)/n]$ to a rational function of type $[n/(n+1)]$. We shall take the poles to be located on the Askey-Wilson grid. It is shown that this operator is related to the one-dimensional degeneration of the Ruijsenaars-van Diejen Hamiltonians.  The Wilson biorthogonal functions of type ${_{10}}\Phi_9$ are found to be solutions of a generalized eigenvalue problem involving rational Heun operators of the special ``classical'' kind.
\end{abstract}

\keywords{}

\maketitle

\begin{center}

{\it Dedicated to the memory of Dick Askey}

\end{center}

\section{Introduction}
The standard Heun operator and its generalizations bear significant interest and have various connections to Painlev\' e equations and integrable models in particular \cite{Takemura_deg, NRY} .  These operators have also arisen in the transfer matrix associated to the reflection algebra and lend themselves to diagonalization via the algebraic Bethe ansatz \cite{BP,KH}. We here identify and discuss what we shall call the rational Heun operator since it is defined through properties with respect to rational functions. We shall observe that it is related to the one-particle degeneration of the Ruijsenaars-van Diejen Hamiltonians \cite{Diejen, Ruijsenaars} studied by Takemura \cite{Takemura_deg}.   The Wilson biorthogonal functions of type ${_{10}}\Phi_9$ will also arise in this framework. 

This continues our investigations on discrete and $q$-difference extensions of the Heun equation. The standard Heun operator defines the second order differential equation with four regular Fuchsian singularities \cite{K, Ronveaux}. It can be characterized as the most general operator of second order in the derivative that maps polynomials of degree $n$ into generic polynomials of degree $n+1$ \cite{VZ_HH}. This operator can also be obtained by applying the so-called tridiagonalization procedure to the hypergeometric operator \cite{GVZ_Heun}. This relies on the bispectrality of the Jacobi polynomials. It is now appreciated that generalizations of the standard Heun operators can be obtained equivalently, either by implementing the degree raising property or by applying the algebraic Heun operator construct or tridiagonalization to some bispectral families of orthogonal polynomials \cite{GVZ_band}. Focusing on the uniform lattice and the Hahn polynomials has led to a discrete version of the Heun operator \cite{VZ_HH}. Looking similarly at the little and big $q$-Jacobi polynomials and their orthogonality grids gave rise to two $q$-versions of the Heun operators \cite{BVZ_Heun} (one actually introduced by Hahn \cite{Hahn_Heun}). The Heun operator of Askey-Wilson type was analogously determined in \cite{BTVZ} and those associated to Lie algebras were found in \cite{CVZ}.

The most general $q$-Heun operator associated in the fashion described above to polynomials on the linear $q$-grid was shown \cite{BVZ_Heun} to coincide with a degeneration of the Ruijsenaars-van Diejen Hamiltonian called $A^{(3)}$ by Takemura in \cite{Takemura_deg}. The special case connected to the little $q$-Jacobi polynomials (instead of the big $q$-Jacobi polynomials) was readily seen to match the degeneration $A^{(4)}$ of Takemura. Two more $q$-Heun operators $A^{(1)}$ and $A^{(2)}$ were identified by Takemura in \cite{Takemura_deg}.
One objective here is to provide for $A^{(1)}$, the most complicated of these operators, a characterization akin to the one achieved for $A^{(3)}$ and $A^{(4)}$. ($A^{(2)}$ will be discussed separately.)
To that end, we shall move outside the realm of orthogonal polynomials; this will leave us with the raising property as the guide in our search. We shall look for generalized $q$-Heun operators defined by the feature that they map any rational function of the type $[(n-1)/n]$ (see section 1 for a precise definition) with poles at $x_0, x_1,..., x_n$ to one of type $[n/(n+1)]$ with one more pole at $x_{n+1}$. The connection with the operator  $A^{(1)}$ will be made when the poles are located on Askey-Wilson grid.
As a note, we shall also observe that the Heun operator of Askey-Wilson type has a connection with the operator $A^{(1)}$.

Attention will be given to special functions that are associated to the rational Heun operator. We shall call such an operator ``classical'' if its action on a rational function of type  $[(n-1)/n]$ gives another function of that type, namely if one pole say at $x_1$ is suppressed while a pole at $x_{n+1}$ is added. The Wilson biorthogonal rational functions of type ${_{10}}\Phi_9$ \cite{GM} will then be shown to solve generalized eigenvalue problems associated to these ``classical'' Heun operators.

The rest of the paper is structured as follows. In Section 2 we make precise the definition of the rational Heun operator through a raising property on a set of elementary rational functions with poles at $x=x_n$. This definition is implemented 
in Section \ref{sec:correspon} when the poles are located at the sites of the Askey-Wilson lattice; this will yield a $q$-difference operator involving 8 parameters ($x_0$ is expressed in terms of these parameters) and acting  ``tridiagonally''. This rational operator $W$ is then compared with $A^{(1)}$; the two operators are seen to differ by an elementary $[0/1]$ rational function with a pole at a prescribed $\t x_0$. This allows for a complete characterization of $A^{(1)}$ as mapping rational functions of type $[(n-1)/n]$ into ones of type $[(n+1)/(n+2)]$. 
In Section \ref{sec:Degeneration}, it is briefly shown as an aside how the Heun operator associated to Askey-Wilson polynomials can be connected to the one-partical Hamiltonian of Ruijsenaars and van Diejen through a limit of the operator $A^{(1)}$ discussed in Section \ref{sec:correspon}. Section \ref{sec:classical_Heun} is concerned with the definition of the rational Heun operators of a special kind that will be called classical and will be central in the two sections that follow. Section \ref{sec:Difference} explains how generalized eigenvalue problems can be posited in terms of rational Heun operators and it is shown in Section \ref{sec:solutions} how the Wilson biorthogonal rational functions arise as solutions.
The paper ends with concluding remarks in Section \ref{sec:concluding}.

\section{Rational $q$-Heun operator}
\setcounter{equation}{0}
Consider the generic second order $q$-difference operator
\be
W= A_1(z) T^+ + A_2(z) T^- + A_0(z) \mathcal{I} \lab{gen_W} \ee
where $T^+ f(z) = f(zq), \: T^- f(z) = f(z/q)$.
Assume that $\psi(z)=R(x)$ is an arbitrary rational function of order $[n/(n+1)]$ in the argument
\be
x=z+z^{-1} \lab{x-z} \ee
with prescribed poles at $x=x_0,x_1,x_2, \dots$ :
\be
R(x) = \frac{\xi_0}{x-x_0} +  \frac{\xi_1}{x-x_1} + \dots +\frac{\xi_n}{x-x_n} = \frac{Q_1(x)}{Q_2(x)}, \lab{expansion_R} \ee
where $Q_1(x)$ and $Q_2(x)$ are polynomials of degrees not exceeding $n$ and $n+1$, respectively. This defines what we mean by a rational function $R_n$ of order $[n/(n+1)]$.
We formulate the following problem:

{\it Find the most general operator $W$ of the form \re{gen_W} that sends any rational function $R_n(x(z))$ of type $[n/(n+1)]$ with poles prescribed to be at $x_0,x_1,x_2, \dots, x_n$ to another rational function of $\t R_{n+1}(x(z))$ type $[(n+1)/(n+2)]$ with poles at $x_0,x_1,x_2, \dots, x_n, x_{n+1}$ and $x(z)$ given by \re{x-z}}.

In other words the question is to find the most general operator W that acts as follows on the rational function $R_n(x)$:

\be
W R_n(x) = \t R_{n+1}(x) = \frac{\eta_0}{x-x_0}  + \frac{\eta_1}{x-x_1} + \dots +\frac{\eta_n}{x-x_n} + \frac{\eta_{n+1}}{x-x_{n+1}} \lab{t_R} \ee
thereby bringing an additional pole at $x=x_{n+1}$.
We shall refer to such $W$ by the expression {\it rational Heun operator}.

In order to find the solution of this problem, we shall proceed in a stepwise fashion. First, the action of the operator $W$ on the elementary rational function $(x-x_0)^{-1}$ should give a rational function  of type [1/2] with poles at $x=x_0$ and $x=x_1$: 
\be
W \left\{\dfrac{1}{x(z)-x_0}\right\} = \frac{A_1(z)}{zq+(zq)^{-1}-x_0} + \frac{A_2(z)}{z/q+(z/q)^{-1}-x_0} + \frac{A_0(z)}{z+(z)^{-1}-x_0} =  r_1(x(z)) \lab{W-0} \ee
where 
\be
r_1(x(z)) = \frac{\xi_{00}}{x(z) -x_0} + \frac{\xi_{01}}{x(z) -x_1} \lab{r_1_def}. \ee 
Second, W applied to the next elementary rational function $(x-x_1)^{-1}$ should yield 
 a rational function of order $[2/3]$:
\be
W \left\{ \frac{1}{x(z) - x_1} \right\} = \frac{A_1(z)}{zq+(zq)^{-1} -x_1} +\frac{A_2(z)}{z/q+(z/q)^{-1} -x_1} + \frac{A_0(z)}{z+(z)^{-1} -x_1} = r_2(x(z)), \lab{W-1} \ee
where
\be
r_2(x)=\frac{\xi_{10}}{x - x_0} + \frac{\xi_{11}}{x - x_1} + \frac{\xi_{12}}{x - x_2} \lab{r_2_def}. \ee
Finally, at the next step we should have
\be
W \left\{ \frac{1}{x(z) - x_2} \right\} = \frac{A_1(z)}{zq+(zq)^{-1} -x_2} +\frac{A_2(z)}{z/q+(z/q)^{-1} -x_2} + \frac{A_0(z)}{z+(z)^{-1} -x_2} = r_3(x(z)), \lab{W-2} \ee
where
\be
r_3(x)=\frac{\xi_{20}}{x - x_0} + \frac{\xi_{21}}{x - x_1} + \frac{\xi_{22}}{x - x_2} + \frac{\xi_{23}}{x - x_3} \lab{r_3_def}. \ee
This determines $W$. Indeed, from the three equations \re{W-0}, \re{W-1} and \re{W-2} we can find the explicit expressions of the functions 
 $A_i(z)$ entering in the formula for $W$:
\be 
A_1(z) = \frac{(x(zq)-x_0)(x(zq)-x_1)(x(zq)-x_2)}{(x(zq)-x(z))(x(zq)-x(z/q))}
\sum_{j=0}^{2}\frac{(x(z)-x_j)(x(z/q)-x_j) }{(x_j-x_{j+1})(x_j-x_{j+2})} r_{j+1}(z), \lab{B1_expl}\ee
\be 
A_2(z) = \frac{(x(z/q)-x_0)(x(z/q)-x_1)(x(z/q)-x_2)}{(x(z/q)-x(z))(x(z/q)-x(zq))}
\sum_{j=0}^{2}\frac{(x(z)-x_j)(x(zq)-x_j) }{(x_j-x_{j+1})(x_j-x_{j+2})} r_{j+1}(z) \lab{B2_expl}\ee
where $x_{3}=x_0$ and $x_{4}=x_1$.
Note that the functions $A_1(z)$ and $A_2(z)$ are related by
\be
A_2(z) = A_1(1/z). \lab{B1-B2} \ee
Indeed, by definition $x(1/z)=x(z)$, and property \re{B1-B2} follows immediately from \re{B1_expl}-\re{B2_expl}.

So far the locations $x_n$ of the poles were arbitrary.
From now on we shall take them to be:
\be
x_n = \alpha q^n + \frac{1}{\alpha q^n}. \lab{x_n_alpha} \ee
In other words, we shall demand that the poles be at the sites of the Askey-Wilson grid
\be
x_n = x(\alpha q^n) \lab{x_n_x_z} \ee
where $x(z)=z+z^{-1}$.
With this choice for the location of the poles, the definition of $W$ in \re{t_R} can be considered as a natural generalization of the Heun-Askey-Wilson operator proposed in \cite{BTVZ} (see Section \ref{sec:Degeneration}) where a similar raising property, namely, $W P_n(x(z)) = \tilde P_{n+1}(x(z))$ was applied to polynomials instead of rational functions.

In order to recognize the structure of the expressions \re{B1_expl} and \re{B2_expl},  we take note of the following algebraic relations:
\ba
\frac { (x(zq) - x_{0} )( x(zq) - x_{1} )( x(zq) - x_{2} ) }
{ (x(qz) - x(z))( x(zq) - x(z/q)) } = \frac{(z/(\alpha q);q)_3 (\alpha q z;q)_3}{q (q;q)_2\, z\, (z^2;q)_2 } \lab{P1_B1} \ea
and
\be
\sum_{j=0}^{2}\frac{(x(z)-x_j)(x(z/q)-x_j)\, r_{j+1}(z)  }{(x_j-x_{j+1})(x_j-x_{j+2})} 
 = -\frac{q^4 \alpha^2 Q_8(z)}{(q;q)_2 (\alpha^2 q;q)_3\, z \, (z/(\alpha q);q)_2 (\alpha z ;q)_4}, \lab{P2_B1} \ee
where $(a;q)_n=(1-a)(1-a q)\cdots(1-a q^{n-1})$ and $Q_8(z)$ is a polynomial of degree 8 given by 
\be
Q_8(z)=\sum_{j=0}^8 \eta_j z^j \qquad (\eta_8=\alpha^2 q^3 \eta_0)
,\lab{polyQ8}\ee
whose coefficients $\eta_j$ are linear combinations of the elements in the set $\Xi = \{\xi_{00}$, $\xi_{01}$, $\xi_{10}$,$ \xi_{11}$, $\xi_{12}$, $\xi_{20}, \xi_{21}, \xi_{22}, \xi_{23}\}$,
\be
\eta_j = \sum_{\xi \in \Xi } d_{\xi}(\alpha,q)\,\xi.\ee

Combining expressions \re{P1_B1} and \re{P2_B1}, we get 
\be
A_1(z) = \kappa \: \frac{ (qz- \alpha) Q_8(z)}{z^2(1-\alpha z)(1-z^2)(1-qz^2)} \lab{B1_Q} \ee
with
\be 
\kappa = \frac{\alpha q^3}{(1-q)^2(1-q^2)^2(1-\alpha^2q )(1-\alpha^2q^2 )(1-\alpha^2q^3 )}. \lab{kappa_1} \ee
We  recall that $A_0$ is given by relation \re{W-0}. We further note that the rational Heun operator can be presented in the form
\be
W=A_1(z) \left(T^+ - \dfrac{x(z)-x_0}{x(qz)-x_0} \mathcal{I}  \right) +A_1(z^{-1}) \left(T^- - \dfrac{x(z)-x_0}{x(z/q)-x_0}\mathcal{I}  \right) + (x(z)-x_0)\, r_1(z) \mathcal{I} \lab{W_red} \ee
where we have used relation \re{W-0}. The form \re{W_red} is convenient in applications to biorthogonal rational functions of classical type (see Section \ref{sec:classical_Heun}).

\begin{pr}
The operator $W$ given by \re{gen_W},\re{W-0},\re{B1_expl} and \re{B2_expl}  sends any rational function $R_n(x(z))$ of type $[n/(n+1)]$ with poles prescribed to be at $x_0,x_1,x_2, \dots, x_n$ to another rational function $\t R_{n+1}(x(z))$ of type $[(n+1)/(n+2)]$ with poles at $x_0,x_1,x_2, \dots, x_n, x_{n+1}$ if the 
the pole locations $x_n$ are given by \re{x_n_alpha}.

\end{pr}
The proof of this proposition is direct. We apply the operator $W$ to the elementary rational function $(x(z) - x_n)^{-1}$ with its only simple pole at $x=x_n$ and then use the relations
\be
x(qz)+x(z/q) = (q+q^{-1}) x(z), \quad x(qz) x(z/q)= x^2(z) +(q-q^{-1})^2. \lab{xx_prop} \ee
It is easily verified that $W (x(z)-x_n)^{-1}$ is a linear combination of elementary rational functions with poles at $x=x_{k}$ where $k=0,1,\dots,n+1$. This proves the proposition.

Let us denote the elementary rational functions by $\chi _n$:
\be
\quad \chi_n(x) = \frac{1}{x-x_n}. \lab{chi_basis} \ee
Then any rational function $\psi_n(x)$ of order $[n/(n+1)]$ with its poles required to be at $x_0,x_1,x_2, \dots, x_n$ can be presented as the linear combination
\be
\psi_n(x) = \sum_{k=0}^n \beta_{nk} \chi_k(x). \lab{psi_basis} \ee 
The main property of the rational $q$-Heun operator 
can be formulated as follows:
\be
W \chi_n(x) = \sum_{k=0}^{n+1} \gamma_{nk} \chi_k(x) \lab{W_chi} \ee
for some coefficients $\gamma_{nk}$. In fact, for the operator $W$ of \re{W_red}, given the specific form of the functions $A_1(z), A_2(z)$, it can be verified for every $n$ that only the 4 coefficients $\gamma_{n,0}$, $\gamma_{n,n-1}$, $\gamma_{n,n}$, $\gamma_{n,n+1}$ can possibly be nonzero in the linear combination \re{W_chi}.

\section{Correspondence with the Ruijsenaars-van Diejen operators}
\label{sec:correspon}

First we consider how many independent parameters are in the operator $W$. We have 9 parameters $\xi_{ik}$ in the definition of the rational functions $r_1(x),r_2(x),r_3(x)$ and the parameter $\alpha$ which enters in definition \re{x_n_alpha} of the pole grid.
Among these, one parameter in $Q_8(z)$ is a common factor. 
Of these 9 independent parameters, note that one - namely $\xi_{00}$ - is an additive constant in the operator $W$. In Takemura's paper \cite{Takemura_deg} the number of parameters is 8 because the additive parameter is not included.
Let us examine this situation in more details.
The coefficients of the eight degree polynomial $Q_8(z)$ in $A_1(z)$ and $A_2(z)$ give us 8 parameters $\eta_0, \eta_1, \ldots, \eta_7$, which are independent linear combinations of the 9 parameters $\xi_{ik}$.
Similarly we obtain five more parameters $\t\eta_0,\t\eta_1,\ldots,\t\eta_4$ from the coefficients of $A_0(z)$:
\be
A_0(z) = 
\t\eta_1 \left(z^2+\frac{1}{z^2}\right)
+ \t\eta_2\left(z+\frac{1}{z}\right) 
+\frac{\t\eta_3 + \t\eta_4 z}{1-q z^2}
+\frac{\t\eta_3 q + \t\eta_4 z}{z^2-q}
+\t\eta_0, \lab{B0_expl}
\ee
where each of $\t\eta_j, j=0,1,\ldots,4$, is a linear combination of the $\xi_{ik}$.
It is then easy to see that $\t\eta_1,\t\eta_2,\t\eta_3,\t\eta_4$ can be presented as linear combinations of the 8 parameters $\eta_j$, while this is not so for the constant term $\t\eta_0$. 
Hence substracting the constant $\t\eta_0$, we obtain the operator $ W_0 = \kappa^{-1}(W-\t\eta_0)$ which has 9 independent parameters $\alpha,\eta_0,\eta_1,\ldots,\eta_7$  including the common factor:
\be
 W_0 = U(z) \,T^{+} + U(z^{-1}) \,T^{-} + V(z), \label{W0}\ee
where 
\ba
&& U(z) = \frac{(qz-\alpha)Q_8(z)}{z^2 (1-\alpha z)(1-z^2)(1-q z^2)},\\
&& V(z) = 
c_1\left(z^2+\frac{1}{z^2}\right)
+c_2\left(z+\frac{1}{z}\right) +\frac{c_3 + c_4 z}{1-q z^2}
+\frac{c_3 q + c_4 z}{z^2-q}
\ea
with $\eta_8 =\alpha^2 q^3 \eta_0$ and
\ba
&& c_1 = \alpha q \eta_0+(\alpha q)^{-1} \eta_8 = q (q+1) \alpha \eta_0,\\
&& c_2 = \alpha q \eta_1+(\alpha q)^{-1} \eta_7,\\
&& c_3 = q^{-1}(1-q)^{-1}(\eta_7 +\eta_5 q + \eta_3 q^2 +\eta_1 q^3),\\
&& c_4 = q^{-1}(1-q)^{-1}(\eta_8 +\eta_6 q + \eta_4 q^2 +\eta_2 q^3 + \eta_0 q^4).
\ea

Now we introduce another parameterization $\varepsilon_1,\varepsilon_2,\ldots,\varepsilon_8$ 
through
\be
\eta_k = (-p)^k \sigma_k \eta_0  \quad (k=1,2,\ldots, 8)
\ee
where $q=p^2$ and 
$\sigma_k$ denotes the elementary symmetric polynomials of degree $k$ in the variables $\varepsilon_1^2,\varepsilon_2^2,\ldots,\varepsilon_8^2$. 
Thus the condition $\eta_8=\alpha^2 q^3 \eta_0$ turns out to be
\be
\alpha = p\, \varepsilon_{1}\varepsilon_{2}\varepsilon_{3}\varepsilon_{4}\varepsilon_{5}\varepsilon_{6}\varepsilon_{7}\varepsilon_{8}. \ee
In this parametrization the functions $A_1 (z)$ and $A_0(z)$ that define $W$ read: 
\ba
&&A_1(z)=
\frac{\kappa\eta_0 p(pz-\prod_{\ell=1}^8\varepsilon_{\ell} )\prod_{j=1}^8(1-\varepsilon_j^2 pz ) }{z^2(1- pz\prod_{\ell=1}^8\varepsilon_{\ell})(1-z^2)(1-p^2z^2)}, \lab{def:A1_eps}
\\
&&A_0(z)=\tilde \eta_0+\kappa\eta_0p^3\bigg\{\frac{\prod_{j=1}^8 (1-\varepsilon_j^2)}{2(1-z p^{-1})(1-(pz)^{-1})}
-\frac{\prod_{j=1}^8 (1+\varepsilon_j^2)}{2(1+z p^{-1})(1+(pz)^{-1})}
\nonumber
\\ 
&& \phantom{\t V(z)=}
-p\prod_{\ell=1}^{8}\varepsilon_{\ell}
\left(\sum_{j=1}^8\left( \varepsilon_j^2 +\varepsilon_j^{-2} \right)
\left(z+z^{-1}\right)
-\left(p+p^{-1}\right)\left(z^2+z^{-2}\right)\right)\Bigg\}. \lab{def:A0_eps}
\ea
The actions of $W=\kappa W_0+\tilde\eta_0$ on the rational function $\chi_n(x)\,(n=0,1,2,\ldots)$ are explicitly given by
\be
W \chi_n(x) 
=
\t\gamma_{n,0} \chi_0(x)
+\t\gamma_{n,1} \chi_{n-1}(x)
+\t\gamma_{n,2} \chi_{n}(x)
+\t\gamma_{n,3} \chi_{n+1}(x)
\lab{exp_form:W_chi}
\ee
where
\ba
&&{\t\gamma}_{n,0}=\dfrac{\kappa p^{2n} \prod_{j=1}^8 (\alpha - p\varepsilon_j^2 ) \eta_0}{\alpha^3(1-\alpha^2 p^{2n-2})(1-p^{2n+2})},\lab{exp_form_1}\\
&&{\t\gamma}_{n,1}=-\dfrac{\kappa(1-p^{2n}) \prod_{j=1}^8 (1-\alpha p^{2n-1} \varepsilon_j^2 ) \eta_0}{\alpha p^{4n-6}(1-\alpha^2 p^{2n-2})(1-\alpha^2 p^{4n-2})(1-\alpha^2 p^{4n})},\lab{exp_form_2}\\
&&{\t\gamma}_{n,2}=\tilde{\eta}_0+ \kappa\alpha p^3 \biggl\{\left(\alpha^2 p^{4n}+\dfrac{1}{\alpha^2 p^{4n}}\right)\left(p+\dfrac{1}{p}\right)
-\left(\alpha p^{2n}+\dfrac{1}{\alpha p^{2n}}\right)\sum_{j=1}^8\left(\varepsilon_j^2+ \frac{1}{\varepsilon_j^2}\right)
\nonumber\\
&& \qquad - \dfrac{p^{2n+1}}{2}\left(
\dfrac{\prod_{j=1}^8(1-\varepsilon_j^2)}{(1-\alpha p^{2n-1})(1-\alpha p^{2n+1})} 
+\dfrac{\prod_{j=1}^8 (1+\varepsilon_j^2)}{(1+\alpha p^{2n-1})(1+\alpha p^{2n+1})} \right)\Biggr\} \eta_0,\lab{exp_form_3}
\\
&&{\t\gamma}_{n,3}=-\dfrac{\kappa (1-\alpha^2 p^{2n}) \prod_{j=1}^{8} (\alpha p^{2n+1} - \varepsilon_j^2) \eta_0}{\alpha^3 p^{4n-2}(1-p^{2n+2})(1-\alpha^2 p^{4n})(1-\alpha^2 p^{4n+2})}. \lab{exp_form_4}
\ea

With an eye to the connection with the operator $A^{(1)}$ of Takemura we make here the following observation. Suppose the diagonal term of $W_0$ is modified as follows
\be
\hat W= W_0 -\frac{\eta_0 p^3 \prod_{j=1}^8 (1-\varepsilon_j^2)}{(1-zp^{-1})(1-(pz)^{-1})}{\mathcal I}\ee
by the addition of a diagonal term so has to change the sign of the first term in the parentheses of \re{def:A0_eps} that gives $A_0$. 
The raising properties of this modified operator $\hat{W}$ can be determined and are found to be:
\ba
\label{newRaisingProp1}
&& \hat{W}\left\{\dfrac{1}{x - x_0}\right\}=
\dfrac{\gamma^{(1)}_{0,-1}}{x - y_0}+\dfrac{\gamma^{(1)}_{0,0}}{x - x_0}+\dfrac{\gamma^{(1)}_{0,1}}{x - x_1}, \\
\label{newRaisingProp2}
&& \hat{W}\left\{\dfrac{1}{x - y_0}\right\}=
\dfrac{\gamma^{(2)}_{0,-1}}{x - x_0}+\dfrac{\gamma^{(2)}_{0,0}}{x - y_0}+\dfrac{\gamma^{(2)}_{1,1}}{x - y_1}, \\
\label{newRaisingProp3}
&& \hat{W}\left\{\dfrac{1}{x - x_1}\right\}=
\dfrac{\gamma^{(1)}_{1,-1}}{x - y_1}
+\dfrac{\gamma^{(1)}_{1,0}}{x - x_0}
+\dfrac{\gamma^{(1)}_{1,1}}{x - x_1}
+\dfrac{\gamma^{(1)}_{1,2}}{x - x_2}, \\
\label{newRaisingProp4}
&& \hat{W}\left\{\dfrac{1}{x - y_1}\right\}=
\dfrac{\gamma^{(2)}_{1,-1}}{x - x_0}
+\dfrac{\gamma^{(2)}_{1,0}}{x - y_0}
+\dfrac{\gamma^{(2)}_{1,1}}{x - y_1}
+\dfrac{\gamma^{(2)}_{1,2}}{x - y_2},
\ea
and 
\ba
\label{newRaisingProp5}
&& \hat{W}\left\{\dfrac{1}{x - x_k}\right\}=
\dfrac{\gamma^{(1)}_{k,-1}}{x - y_0}
+\dfrac{\gamma^{(1)}_{k,0}}{x - x_0}
+\dfrac{\gamma^{(1)}_{k,k-1}}{x - x_{k-1}}
+\dfrac{\gamma^{(1)}_{k,k}}{x - x_{k}}
+\dfrac{\gamma^{(1)}_{k,k+1}}{x - x_{k+1}}, \\
&& \hat{W}\left\{\dfrac{1}{x - y_k}\right\}=
\dfrac{\gamma^{(2)}_{k,-1}}{x - x_0}+\dfrac{\gamma^{(2)}_{k,0}}{x - y_0}
+\dfrac{\gamma^{(2)}_{k,k-1}}{x - y_{k-1}}
+\dfrac{\gamma^{(2)}_{k,k}}{x - y_{k}}
+\dfrac{\gamma^{(2)}_{k,k+1}}{x - y_{k+1}}, 
\label{newRaisingProp6}
\ea
for $k\ge 2$, where $x_n = p^{2n+1} \prod_{\ell=1}^8 \epsilon_{\ell}+p^{-2n-1} \prod_{\ell=1}^8 \epsilon_{\ell}^{-1}$ and $y_n = p^{2n+1} + p^{-2n-1}$.
We thus see that this modification of $W_0$ bring two series of poles in play.

We are now ready to discuss the relationship between 
the rational Heun operator with the simple raising property involving only one series of poles at the points of the Askey-Wilson grid and the $BC_1$ Ruijsenaars-van Diejen Hamiltonian.
By using the gauge function 
\be
\Psi(z) = (p \alpha z;p^2)_{\infty}(p \alpha z^{-1};p^2)_{\infty}(p z;p^2)_{\infty}(p z^{-1};p^2)_{\infty}\ee
that satisfies
$\Psi(p^2 z) =-\dfrac{1-\alpha z^{-1}p^{-2}}{p z(1-\alpha z)} \Psi(z)$,
we obtain Takemura's $A^{(1)}$ operator:
\be
A^{(1)}=-\eta_0^{-1}p^{-3}\Psi(z) \hat W \Psi(z)^{-1} = \t U(z) \,T^{+} + \t U(z^{-1}) \,T^{-} + \t V(z), \label{tW0}
\ee
where 
\ba
&&\t U\left(z\right)=
-\frac{\Psi(z)U(z)}{\Psi(p^2z)\eta_0 p^{3}} = \frac{Q_8(z)/\eta_0}{(1-z^2)(1-q z^2) } = \frac{\prod_{j=1}^8(1-\varepsilon_j^2 pz ) }{(1-z^2)(1-p^2z^2)},\\
&& \t V(z)=
\frac{\prod_{j=1}^8 (1-\varepsilon_j^2)}{2(1-z p^{-1})(1-(pz)^{-1})}
+\frac{\prod_{j=1}^8 (1+\varepsilon_j^2)}{2(1+z p^{-1})(1+(pz)^{-1})}
\nonumber
\\ 
&& \phantom{\t V(z)=}
+p\prod_{\ell=1}^{8}\varepsilon_{\ell}
\left(\sum_{j=1}^8\left( \varepsilon_j^2 +\varepsilon_j^{-2} \right)
\left(z+z^{-1}\right)
-\left(p+p^{-1}\right)\left(z^2+z^{-2}\right)\right).\ea
We thus observe that the operator $A^{(1)}$ is characterized as per (\ref{newRaisingProp1})-(\ref{newRaisingProp6}) with  a raising property involving two series of poles.
It is however simply obtained from the rational Heun operator $W_0$ by the addition of a diagonal term.

\section{The Askey-Wilson Heun operator and a new degeneration of the Ruijsenaars-van Diejen operators}\label{sec:Degeneration}

The Heun operator associated to the Askey-Wilson polynomials was constructed in \cite{BTVZ} where it was obtained in two equivalent ways.
On the one hand, it was written down by applying the algebraic Heun operator construct or tri-diagonalization method to the bispectral operators of the Askey-Wilson polynomials. On the other hand, it was arrived at by enforcing a raising property. In the latter approach, one looked at the most general $q$-difference operator of second order that maps polynomials of degree $n$ on the Askey-Wilson grid to polynomials of the degree $n+1$ on that same grid.
In both ways one finds the operator 
\be W_{\!\text{AW}} = A_{1}^{(\text{AW})}(z) \left(T^{+}-{\mathcal I}\right) +  A_{2}^{(\text{AW})}(z) \left(T^{-}-{\mathcal I}\right) + p_1(x){\mathcal I}
\ee
with 
\be A_{1}^{(\text{AW})}(z) =\dfrac{Q_6(z)}{z(1-z^2)(1-p^2z^2)}, \quad  A_{2}^{(\text{AW})}(z) =  A_{2}^{(\text{AW})}\left(1/z\right)\ee
where $Q_6$ is an arbitrary polynomial of degree 6 and $p_1(x)$ a polynomial of degree one in $x=z+1/z$.

We here indicate that $W_{\!\text{AW}}$ can be recovered from the degeneration $A^{(1)}$ of the Ruijsenaars - van Diejen Hamiltonian through a limiting procedure.
Let $\varepsilon_7 = \delta_7 t$ and $\varepsilon_8 = \delta_8 t^{-1}$.
Under the limit $t \to \infty$, the operator 
\be
-\delta_7^{-2} t^{-2}\left( p^{-2}z^{-2}\t U(z) T^{+}
+ p^{-2}z^{2}\t U(z^{-1}) T^{-}
+\t V(z) {\mathcal I}
\right)
\ee
reduces to
\be
\hat U(z) (T^{+}-{\mathcal I})+\hat U(z^{-1}) (T^{-}-{\mathcal I}) + \hat V(z){\mathcal I}
\ee
where
\ba
&&\hat U(z)=\dfrac{(1-\varepsilon_1^2 p z)(1-\varepsilon_2^2 p z)(1-\varepsilon_3^2 p z)(1-\varepsilon_4^2 p z)(1-\varepsilon_5^2 p z)(1-\varepsilon_6^2 p z)}{p z (1-z^2)(1-p^2 z^2 )},\\
&& \hat V(z)= \dfrac{(p^2\varepsilon_1\varepsilon_2\varepsilon_3\varepsilon_4\varepsilon_5\varepsilon_6 \delta_7 \delta_8  -1 )(p^2\varepsilon_1\varepsilon_2\varepsilon_3\varepsilon_4\varepsilon_5\varepsilon_6  -\delta_7\delta_8)}{p \,\delta_7\delta_8}\left(z+z^{-1}\right) + (const).
\ea
One thus recognizes that the resulting operator coincides with $W_{\!\text{AW}}$.

\section{Classical Heun operators}\label{sec:classical_Heun}
\setcounter{equation}{0}
In the following we wish to discuss certain difference equations involving rational Heun operators. Special operators that will be called classical will play a central role in this respect. This section is dedicated to their definition and characterization.

We shall focus on operators $W$ for which 
 $\gamma_{n,1}=0$ in the linear combination \re{W_chi},
that is operators with actions
\ba
&&{\mathcal W}\left\{\frac{1}{x-x_0}\right\} = \frac{\gamma_{00}}{x-x_0}, \quad 
{\mathcal W}\left\{\frac{1}{x-x_1}\right\} = \frac{\gamma_{10}}{x-x_0} + \frac{\gamma_{12}}{x-x_2}, \quad \nonumber\\
&&{\mathcal W} \left\{\frac{1}{x-x_2}\right\} = \frac{\gamma_{20}}{x-x_0} + \frac{\gamma_{22}}{x-x_2} + \frac{\gamma_{23}}{x-x_3}, 
\dots, 
 \lab{W_chi_cl0} \ea
such that the pole at $x=x_1$ is absent from the transformation of the elementary rational functions.

This means that rational Heun operators ${\mathcal W}$ of this type send any rational function of order $[n/(n+1)]$ with poles located at $x_0, x_1,x_2, \dots, x_n$ to a rational function of {\it the same} order $[n/(n+1)]$ with poles at $x_0,x_2,x_3, \dots, x_{n+1}$:
\be
{\mathcal W} \psi_n(x) = \tilde \psi_n(x). \lab{psi-tpsi} \ee
In other words, under the action of $W$ on the function $\psi_n$ with poles at $x_0$ and $x_n, \: n=1,2,\dots$, the pole at $x_0$ stays there and the others are shifted to $x_{n+1}, n=1,2,\dots$.
In this case the expression for the function $A_1(z)$ becomes simpler:
\be
A_1(z) =  \kappa_2 \: \frac{ (qz-\alpha)(z-\alpha q)(\alpha q z-1)Q_6(z)}{z^2(1-\alpha z)(1-z^2)(1-q z^2 )} \lab{A1_Q_cl} \ee
with 
$Q_6(z)$ a polynomial of degree 6 in $z$ and $\kappa_2$ the appropriate constant.
The number of independent parameters in classical Heun operators is indeed 6 as we need to set $\xi_{01}, \xi_{11}$ and $\xi_{21}$ equal to zero in the expressions of the rational functions $r_1(x),r_2(x),r_3(x)$.

Next we  apply the gauge transformation 
${\mathcal W} \to  \hat {\mathcal W} =-{\alpha}{\kappa^{-1} \eta_0^{-1}} \,(x(z)-x_0) {\mathcal W} (x(z)-x_0)^{-1}$
to obtain
\ba
&& \hat {\mathcal W} = B_1(z) (T^+ -{\mathcal I})+B_1(z^{-1}) (T^- -{\mathcal I})+\hat\gamma_{00}{\mathcal I}\lab{classical Heun}
\ea
where 
\be
 B_1(z) = \dfrac{(z-\alpha)(z-\alpha q) Q_6(z)}{z^2 (1-z^2)(1-qz^2)}.
\ee
This transformation removes the pole at $x_0$, but preserves the rest of the poles. 
Namely the operator $\hat {\mathcal W}$  sends any rational function of order $[n/n]$ with poles at $x_1,x_2, \ldots,x_n$ to a rational function of the same order $[n/n]$ with the poles shifted to $x_2,x_3, \ldots, x_{n+1}$:
\be
\hat {\mathcal W}\{1\} = \hat\gamma_{00},  \hat {\mathcal W} \left\{\frac{1}{x-x_1}\right\} = \hat\gamma_{10} + \frac{\hat\gamma_{12}}{x-x_2}, \dots, \hat {\mathcal W} \left\{\frac{1}{x-x_n}\right\} = \hat\gamma_{n0} + \frac{\hat\gamma_{n2}}{x-x_2} + \dots +  \frac{\hat\gamma_{n,n+1}}{x-x_{n+1}}. \lab{W_chi_cl} \ee
The upshot is that the pole at $x=x_1$ is absent in the expansions.
Rational $q$-Heun operators mapping rational functions of degree $[n/n]$ onto themselves and with this particular property concerning the pole at $x_1$ will be called \textit{classical}.

Consider two different classical Heun operators $\hat{\mathcal W}_1$ and $\hat {\mathcal W}_2$. 
They will differ by the choices of the rational functions $r_1(x), r_2(x), r_3(x)$. 
Consider a third operator
\be
\hat{\mathcal W}_3 = \tau_1 \hat {\mathcal W}_1 + \tau_2 \hat{\mathcal W}_2 \lab{W_3} \ee
with arbitrary coefficients $\tau_1, \tau_2$. $\hat{\mathcal W}_3$ 
obviously is also classical and this allows for a possible standardization. Since the leading and the constant coefficients in the polynomial $Q_6$ are proportional (recall that $Q_8$ from which $Q_6$ derives has this property - see \re{polyQ8}), these two terms can be eliminated in $\hat W_3$ by an appropriate choice of $\tau _1$ and $\tau _2$.
In this case the function 
$B_1(z)$ of the operator $\hat{\mathcal W}_3$ will become
\be
B_1(z) =  \kappa_3 \: \frac{ (z-\alpha)(z-\alpha q)Q_4(z)}{z(1-z^2)(1-qz^2)} \lab{B1_Q_min} \ee 
with $Q_4(z)$ a polynomial of degree 4 and $\kappa_3$ again a constant.
Classical rational Heun operators brought to that form will be said to be \textit{minimal}.

When the parameters $\varepsilon_1, \varepsilon_2, \ldots, \varepsilon_8$ are used, the expansion coefficients $\gamma_{n,j}$ are explicitly given in (\ref{exp_form:W_chi})-(\ref{exp_form_4}). It follows that the conditions $\gamma_{n,1}=0 \: (n=0,1,2,\ldots)$ hold 
if we take for instance $\varepsilon_7^2 = \alpha p, \varepsilon_8^2 = \alpha^{-1} p^{-3}$ or $p\,\varepsilon_7\varepsilon_8=p\,\varepsilon_1\varepsilon_2\varepsilon_3\varepsilon_4\varepsilon_5\varepsilon_6\varepsilon_{7}^{-2}=1$, and solve (\ref{exp_form_3}) for $n=1$ to determine $\tilde\eta_0$. 
Hereafter for simplicity, we shall write $\varepsilon_{jk\cdots {\ell}} = \varepsilon_j \varepsilon_k \cdots \varepsilon_{\ell}$ for products of $\varepsilon_j$.
Under this particular choice of parameters, one has the following detailed formulas for the corresponding rational Heun operator:
\be
{\mathcal W} = A_1(z) \left(T^{+} - \dfrac{x(z)-x_0}{x(qz)-x_0} \mathcal{I}\right)
+A_1(z^{-1}) \left(T^{-} - \dfrac{x(z)-x_0}{x(z/q)-x_0}\mathcal{I} \right)
 + \gamma_{00}\mathcal{I}
\lab{W with skip pole}
\ee
where 
\ba
&& A_1(z) = -\kappa\eta_0\dfrac{(\varepsilon_{123456} - p^2 z )(\varepsilon_{123456} p^2 - z)(1-\varepsilon_{123456} p^2 z) \prod_{j=1}^{6}(1-\varepsilon_j^2 pz) }{\varepsilon_{123456} \,p^2 z^2(1-\varepsilon_{123456} z)(1-z^2)(1-p^2 z^2)},\\ 
&&  \gamma_{00} = -\dfrac{\kappa \eta_0}{(\varepsilon_{123456})^{3}} \prod_{j=1}^{6} (\varepsilon_{123456} \,p - \varepsilon_j^2) 
\ea
with $\varepsilon_{123456} =\alpha =p\varepsilon_{12345678}$.
The classical Heun operator $\hat {\mathcal W}$ \re{classical Heun} is then explicitly written as
\be
 \hat {\mathcal W} 
= B_1(z) \left(T^{+} - \mathcal{I}\right)
+B_1(z^{-1}) \left(T^{-} - \mathcal{I} \right)
 + \hat\gamma_{00}\mathcal{I}
\ee
where
\ba
&&B_1(z)
 = \dfrac{(z-\varepsilon_{123456})(z-\varepsilon_{123456}\, p^2)}
{z^2(1-z^2)(1-p^2z^2)}\prod_{j=1}^6(1- \varepsilon_j^2 pz),\\
&& \hat\gamma_{00} = \prod_{j=1}^{6} \left(1 - \varepsilon_{123456} \varepsilon_j^{-2} p \right)
\ea
and we observe as should be that
 the classical Heun operator $\hat {\mathcal W}$ contains 6 independent constants $\varepsilon_1, \varepsilon_2, \varepsilon_3, \varepsilon_4,\varepsilon_5,\varepsilon_6$ and $p=q^{1/2}$.

\section{Difference equations with rational Heun operators}\lab{sec:Difference}
We are now equipped to discuss $q$-difference equations that can be defined using rational Heun operators. The presentation will consist of two parts. In the first we shall indicate how one is led to generalized eigenvalue problems (GEVP) involving classical operators. In the second, much shorter, we shall explain how an eigenvalue problem for rational Heun operator can be posited upon imposing truncation conditions.

\subsection{Generalized eigenvalue problems}
Consider the difference equation
\be
W \psi(x(z)) =0 \lab{W_psi_0} \ee
where $W$ is a rational Heun operator and $\psi(x)$ is a rational function of order $[n/n]$ with poles prescribed to be at $x_1, x_2, \dots , x_n$.

Using the elementary rational basis $\chi_n(x)$, we have the expansion \re{W_chi} with the coefficients $\gamma_{nk}$. We can also expand the function $\psi(x)$ over the same basis \re{psi_basis} with some coefficients $\beta_{nk}$. Equation \re{W_psi_0} is then equivalent to the following set of algebraic equations
\be
\sum_{i=0}^n \beta_{ni} \gamma_{ik} =0, \quad k=0,1,\dots, n+1 \lab{beta_gamma_eq} \ee
for the unknowns $\beta_{ni}$. We thus have the $n+2$ equations \re{beta_gamma_eq} for $n+1$ unknowns $\beta_{ni}, \: i=0,1,\dots, n$, which means that the system \re{beta_gamma_eq} is in general incompatible.

This problem will be straightened if we can reduce the
number of equations to $n+1$. From the previous section we
already know that such a reduction happens for classical Heun
operators because in this case, for every $n$, we have no more than $n+1$ nonzero
expansion coefficients: $\gamma_{n0}, \gamma_{n2}, \gamma_{n3}, \dots,
\gamma_{n,n+1}$. 
It then follows in this instance that the eigenvalue equation \re{W_psi_0}
has a unique solution (apart from some singular exceptional
cases). Hence to proceed with \re{W_psi_0}, we need to
restrict to classical Heun operators.

Classical Heun operator $W$ contains 6 arbitrary parameters including the free additive one. If we take this parameter as the eigenvalue $\lambda$, we then arrive at the ordinary eigenvalue problem
\be
W \psi(x) = \lambda \psi(x). \lab{ordinary_EVP} \ee
However, this problem cannot have  solutions. Indeed, the lhs of \re{ordinary_EVP} has a pole at $x=x_{n+1}$ while the rhs of \re{ordinary_EVP} does not.

We must hence look for more elaborate eigenvalue problems, namely multiparameter problems \cite{Atk}, \cite{Sleeman}. The simplest such problem is the generalized eigenvalue problem (GEVP) of the type
\be
W_1 \psi = \lambda W_2 \psi, \lab{GEVP} \ee
with two classical Heun operators $W_1$ and $W_2$.
From the above considerations, it follows that the GEVP \re{GEVP} almost always has a unique solution.

We thus see that starting from a pair of classical Heun operators we can construct a solution to a GEVP for a rational function $\psi(x)$. It is clear that instead of \re{GEVP} we can consider the problem
\be
\t W_1 \psi(x) = \t \lambda \t W_2 \psi \lab{t_GEVP} \ee
with the same eigenfunction $\psi(x)$ 
and the linear combinations
\be
\t W_1 = \tau_1 W_1 + \tau_2 W_2, \quad \t W_2 = \rho_1 W_1 + \rho_2 W_2 \lab{t_W} \ee
where $\tau_i, \rho_i$ are arbitrary parameters with the obvious restriction $\tau_1 \rho_2 - \tau_2 \rho_1 \ne 0$.

The new eigenvalue is
\be
\t \lambda = \frac{\tau_1 \lambda + \tau_2}{\rho_1 \lambda+\rho_2}. \lab{t_lambda} \ee
From the previous section we also know that starting with two classical Heun operators $W_1, W_2$ we can construct a ``minimal'' classical Heun operator $W_{\text{min}}$ by an appropriate choice of the parameters $\tau_1, \tau_2$. This means that for the GEVP \re{t_GEVP} we can always take one of these operators as $W_{\text{min}}$, reducing the GEVP to 
\be
W_1 \psi = \lambda W_{\text{min}} \psi. \lab{GEVP_min} \ee 
In the next section we show that the GEVP of the Wilson biorthogonal rational functions is of that type.

\subsection{Finite dimensional reductions}
Apart from considering classical Heun operators in GEVP, we can also set up  ordinary eigenvalue problems with regular rational Heun operator provided, as we shall explain, restrictions are imposed.

Assume that in expansion \re{W_chi} we have the condition  $\gamma_{N,N+1}=0$ for some $N=1,2,\dots$. Then the $N+1$-dimensional space of rational functions with poles at $x=x_0, x_1, x_2, \dots, x_N$ is invariant under action of the Heun operator $W$. Hence, by elementary linear algebra, the operator $W$ becomes a matrix of size $(N+1)\times (N+1)$ acting on this space. This operator has in general $N+1$ eigenvalues $\lambda_n$ associated to the eigenfunctions $\psi_n(x)$ 
\be
W \psi_n(x) =\lambda_n \psi_n(x) , \quad n=0,1,\dots, N. \lab{W_psi_k} \ee
We should stress however, that {\it all} rational functions in \re{W_psi_k} 
have poles at all the sites
$x=x_0$, $x_1$, $x_2, \dots, x_N$.
From \re{exp_form_4}, one can see that $\varepsilon_1^2 = \alpha p^{2N+1}$ for instance will complete the condition $\gamma_{N,N+1}=0$; the matrix representation of $W$ with respect to the basis $1/(x-x_0),1/(x-x_1),\ldots,1/(x-x_N)$ is in this case given by 
\be
M=\left(\delta_{0,j}\tilde\gamma_{i,1}+\delta_{i,j+1}\tilde\gamma_{i,2}+\delta_{i,j}\tilde\gamma_{i,3}+\delta_{i+1,j}\tilde\gamma_{i,4}\right)_{0\le i,j \le N}.
\ee
Let $v_n={}^{t}(v_{n,0},\ldots,v_{n,N})$  for $n=0,1,\ldots,N$ be the eigenvector of $M$ associated with the eigenvalue $\lambda_n$.
The rational eigenfunctions of \re{W_psi_k} will then read:
\be
\psi_n(x)= \sum_{j=0}^{N}\dfrac{v_{n,j}}{x-x_j}
\ee

This can be considered as a generalization of the well-known finite-dimensional (polynomial) reduction of the ordinary Heun operator \cite{Tur_quasi}.

\section{Explicit solutions for GEVP}\label{sec:solutions}
We shall carry in the following the explicit characterization of solutions to GEVPs of the type introduced in the last Section. 
First, we shall consider a situation where a given classical Heun operator with an additive parameter is split in two parts to yield a GEVP. Second we shall explain the occurrence as solutions of the Wilson biorthogonal functions. Third, we shall discuss  how and when can rational solutions to GEVPs defined by two classical Heun operators can be explicitly constructed. 

\subsection{Splitting situation}
Let us return to the classical Heun operator $\hat W$ given in (5.12) - (5.14).
Define
\ba
 && \hat B(z) = \dfrac{p \varepsilon_5\varepsilon_6(z-\varepsilon_{123456})(z- \varepsilon_{123456}\,p^2)\prod_{j=1}^4(1-\varepsilon_j^2 p z)}{z (1-z^2)(1-p^2z^2)} .
\ea
The equation $\hat{\mathcal W} f(z)=0$ can be rewritten into the generalized eigenvalue problem form
\be
 \hat{\mathcal W}_1 f(z) = \lambda \hat{\mathcal W}_2 f(z) 
\lab{GEVP:W1W2}\ee
where
\ba
 && \hat{\mathcal W}_1 = x(\varepsilon_5\varepsilon_6 pz) \hat B(z)  (T^+-{\mathcal I}) +   x(\varepsilon_5\varepsilon_6 pz^{-1}) \hat B(z^{-1})(T^--{\mathcal I}) \nonumber\\
&&\qquad\qquad +   x(\varepsilon_{1234}\,p) \varepsilon_{1234}\,p  \prod_{j=1}^4 (1 - \varepsilon_{123456} \varepsilon_j^{-2}\,p ){\mathcal I},\\
 && \hat{\mathcal W}_2 = \hat B(z) (T^+-{\mathcal I}) + \hat B(z^{-1}) (T^--{\mathcal I}) +  
\varepsilon_{1234}\,p  \prod_{j=1}^4 (1 - \varepsilon_{123456} \varepsilon_j^{-2}\,p ){\mathcal I}
\ea
and
\ba
 && \lambda =  x\left(\dfrac{\varepsilon_5}{\varepsilon_6}\right)
= \dfrac{\varepsilon_5}{\varepsilon_6}
+\dfrac{\varepsilon_6}{\varepsilon_5}.
\ea
Both operators $\hat{\mathcal W}_1$ and $\hat{\mathcal W}_2$ do not depend on the parameter $\varepsilon_5/\varepsilon_6$ which plays the role of the eigenvalue.

It is readily seen that  $\hat{\mathcal W}_1$ and $\hat{\mathcal W}_2$ belong to the family of classical rational Heun operators that $\hat{\mathcal W}_2$ is ``minimal'' and that $\hat{\mathcal W}_1$ is a general classical rational $q$-Heun operator.

Consider at this point the following rational basis of type $[n/n]$
\be 
 \omega_n(x(z))=\breve\omega_n(z;\varepsilon_1^2 p,\varepsilon_{123456} p^2) 
\lab{2diagonal_basis}\ee
where 
\be
\breve\omega_n(z;a,b)  =\dfrac{(a z;p^2)_n(a z^{-1};p^2)_n}{(b z;p^2)_n(b z^{-1};p^2)_n}
\ee and
$(a;q)_n = (1-a)(1-aq)\cdots(1-aq^{n-1})$.
It is directly verified that both operators $\hat{\mathcal W}_1$ and $\hat{\mathcal W}_2$ are two-diagonal in this basis
\ba
&& \hat{\mathcal W}_1\omega_n(x)=\mu_{1,n}\omega_n^{+}(x)+ \nu_{1,n}\omega_{n-1}^{+}(x),\\
&& \hat{\mathcal W}_2\omega_n(x)=\mu_{2,n}\omega_n^{+}(x)+ \nu_{2,n}\omega_{n-1}^{+}(x)
\ea
where 
\ba
\lab{2diag:coe1}
&& \mu_{1,n}=
\left(\varepsilon_{1234}p^{2n+1}  + (\varepsilon_{1234})^{-1}p^{-2n-1} \right)
 \mu_{2,n},\\
&& \nu_{1,n}=
\left(\varepsilon_1^2\varepsilon_5\varepsilon_6 p^{2n} 
+ (\varepsilon_1^2\varepsilon_5\varepsilon_6)^{-1}p^{-2n}  \right)\nu_{2,n},\\
&& \mu_{2,n}=\dfrac{p\,(\varepsilon_1^2 \varepsilon_{123456}p^{2n+1} -1) (\varepsilon_{123456}p - \varepsilon_1^2)\prod_{j=2}^{4}(\varepsilon_{123456}p^{2n+1} - \varepsilon_j^2)}{\varepsilon_{1234} (\varepsilon_1^2 \varepsilon_{123456}p^{4n+1} -1)},\\
&& \nu_{2,n}=
\dfrac{(1 - p^{2n})(\varepsilon_{123456} p - \varepsilon_1^2)\varepsilon_5\varepsilon_6 \prod_{j=2}^{4}(\varepsilon_1^2\varepsilon_j^2 p^{2n} - 1)}{\varepsilon_1^2 \varepsilon_{123456} p^{4n+1} -1},
\lab{2diag:coe4}
\ea
and with the basis 
\be\omega_n^{+}(x)=\breve\omega_n(z;\varepsilon_1^2 p,\varepsilon_{123456} p^4)\ee
obtained from the basis $\breve\omega$ under the substitution $(\varepsilon_1,\varepsilon_{123456}) \to (\varepsilon_1,\varepsilon_{123456}p^2 )$ of the parameters.

Using this basis \re{2diagonal_basis}, we can explicitly construct the rational eigensolutions of the GEVP \re{GEVP:W1W2}.  Let us expand the rational function $R_n(x)$ of type $[n/n]$ over the $\omega_n(x)$:
\be
 R_n(x)=\sum_{s=0}^n A_{ns} \omega_s(x).
\ee
From the GEVP \be\hat{\mathcal W}_1R_n(x) = \lambda_n \hat{\mathcal W}_2R_n(x),\lab{GEVP_W1W2}\ee we get the recurrence relation for the coefficients $A_{ns}$:
\be
 A_{n,s+1}= A_{n,s} \dfrac{\lambda_n \mu_{2,s} - \mu_{1,s}}{\nu_{1,s+1} - \lambda_n \nu_{2,s+1}} 
\ee
where \be\lambda_n = \dfrac{\mu_{1,n}}{\mu_{2,n}}.\ee
This yields the
following explicit formula for the rational function 
\be
 R_n(x) = A_{n,0}\sum_{s=0}^n
\prod_{j=0}^{s-1} \dfrac{\mu_{1,n} \mu_{2,j}- \mu_{1,j}\mu_{2,n}}{\nu_{1,j+1}\mu_{2,n} - \mu_{1,n} \nu_{2,j+1}} \omega_s(x).\lab{HGtypeRAT1}
\ee

\subsection{Wilson biorthogonal rational functions and GEVP}
Let us introduce the function \cite{GM}
\be
\phi = {_{10}}W_9(a;b,c,d,e,f,g,h;q) = {_{10}}\Phi_9 \left({a^2, q {a} , -q {a}, b,c,d,e,f,g,h \atop  {a}, -{a}, \frac{a^2 q}{b}, \frac{a^2 q}{c}, \frac{a^2 q}{d}, \frac{a^2 q}{e}, \frac{a^2 q}{f}, \frac{a^2 q}{g}, \frac{a^2 q}{h}  }  ; q\right). \lab{phi_def} \ee
We assume that the hypergeometric function is very-well-poised. This means that
\be 
bcdefgh = a^6 q^2. \lab{vwp} \ee
Moreover we suppose also that this function is truncated, i.e. that 
we have
\be
g=q^{-n}, \; h= \mu q^n, \quad n=0,1,2,\dots \lab{tr_c} \ee
where the parameter $\mu$ should satisfy the condition \re{vwp}:
\be
\mu bcdef = a^6 q^2. \lab{vwp1} \ee
We can obtain a set of biorthogonal rational functions in the argument $z$ if one takes the following dependence of the parameters
\be
b= \kappa z, \; c = \kappa z^{-1}. \lab{hg_z} \ee
Indeed, it is easily seen that in this case the function $\phi_n(z)=R_n(x)$ becomes a rational function $R_n(x)$ of type $[n/n]$ in the argument
\be
x=z+z^{-1}\lab{x_z} \ee
 with prescribed poles located at the points
\be
x_s= \frac{a^2 q^{s}}{\kappa} + \frac{\kappa}{a^2 q^{s}}, \quad s=1,2, \dots, n \lab{poles_zs} \ee
More explicitly one has
\be
R_n(x) = \xi_{n0} + \frac{\xi_{n1}}{x-x_1} + \dots + \frac{\xi_{nn}}{x-x_n} \lab{R_n_expansion} \ee
with some coefficients $\xi_{ns}$.

Substituting \re{2diag:coe1}-\re{2diag:coe4} into \re{HGtypeRAT1}, the rational function $R_n(x)$ turns out to be the very-well-poised balanced hypergeometric function
\be
 R_n(x(z)) = \mbox{}_{10}B_9\left(\varepsilon_1 \sqrt{ \varepsilon_{123456} p};(\varepsilon_1)^2 p z,\dfrac{(\varepsilon_1)^2 p}{z},\dfrac{\varepsilon_{123456} p}{(\varepsilon_2)^2},\dfrac{\varepsilon_{123456} p}{(\varepsilon_3)^2},\dfrac{\varepsilon_{123456} p}{(\varepsilon_4)^2},p^{-2n};p^2\right),\\
\ee
where 
\be
\mbox{}_{10}B_9\left(a;b,c,d,e,f,g;q\right)=\mbox{}_{10}W_9\left(a;b,c,d,e,f,g,\dfrac{a^6 q^2}{bcdefg};q\right),
\ee
which gives a rational eigensolution of the GEVP \re{GEVP_W1W2}.

Taking $\lambda = \lambda_n$, we obtain 
\be\dfrac{\mu_{1,n}}{\mu_{2,n}} = \dfrac{\varepsilon_5}{\varepsilon_6}+\dfrac{\varepsilon_6}{\varepsilon_5}.\ee
Hence when $\varepsilon_5 (\varepsilon_6)^{-1} = \varepsilon_{1234} p^{2n+1}$ or $(\varepsilon_{1234})^{-1} p^{-2n-1}$,
the Wilson biorthogonal rational function $R_n(x)$ becomes  a solution of
\be\hat{\mathcal W} R_n(x(z))=0\ee
with $\hat{\mathcal W}$ the generic rational Heun operator of classical type\cite{Spiridonov}.

It is directly verified that the action of the operator $\hat{\mathcal W}_1$ (as well as the action of the operator $\hat{\mathcal W}_2$) on the function $R_n(z)$ transforms this function to another Wilson biorthogonal function with shifted parameters 
\be
\hat{\mathcal W}_1 R_n(x) = const \: R_n^{+}(x), \lab{W_1_R} \ee
where $R_n^{+}(x)$ is obtained from $R_n(x)$ by shifting $\varepsilon_{123456} \to \varepsilon_{123456}p^2 $.

Relation \re{W_1_R} is in agreement with the main property of the classical rational $q$-Heun operators: they transform any rational function of order $[n/n]$ with poles at given locations $x_1,x_2, \dots, x_n$ to another rational function of the same order $[n/n]$ with poles displaced to $x_2, x_3, \dots, x_{n+1}$. 

\subsection{Two classical Heun operators}
Finally, let us consider the GEVP corresponding to a pair of generic classical Heun operators $\hat{\mathcal W}$ and $\hat{\mathcal Y}$:
\be
 \hat{\mathcal W} R_n(x) = \lambda_n \hat{\mathcal Y} R_n(x).\lab{GEVP:WY}
\ee
Remarkably, the rational basis $\omega_n(x)$ \re{2diagonal_basis} 
happens to be the two-diagonal basis of the generic classical Heun operator ${\mathcal W}$ defined in \re{classical Heun}:
\be
 \hat{\mathcal W}\omega_n(x)=\mu_{n}\omega_n(x)+ \nu_{n}\omega_{n-1}(x),
\ee
where
\ba
\lab{2diagW:coe1}
&& \mu_{n}=
\dfrac{(\varepsilon_{1234} \varepsilon_6 p^{2n+1} - \varepsilon_5)(\varepsilon_{1234} \varepsilon_5 p^{2n+1} - \varepsilon_6)}{\varepsilon_{123456}  p^{2n+1}  } 
\mu_{2,n},\\
\lab{2diagW:coe2}
&& \nu_{n}=
\dfrac{( (\varepsilon_1\varepsilon_5)^2p^{2n} - 1)((\varepsilon_1\varepsilon_6)^2 p^{2n} -1)}{(\varepsilon_1)^2 \varepsilon_5\varepsilon_6 p^{2n}}
\nu_{2,n}.
\ea
If we take the operator $\hat{\mathcal Y}$ as the operator obtained from $\hat{\mathcal W}$
given in  \re{classical Heun} under the replacement of the six parameters $\varepsilon_j$ by $\delta_j$ with
 the conditions $\varepsilon_1 = \delta_1, \varepsilon_{23456}=\delta_{23456}$, then it is found that  $\hat{\mathcal Y}$ has the same two-diagonal basis as $\hat{\mathcal W}$. This thus shows that the rational eigensolutions of the GEVP defined by two generic classical Heun operators can be explicitly constructed.

\vspace{5mm}

\section{Concluding remarks}\label{sec:concluding}

By considering a raising property on rational functions with poles residing on the grid points of the Askey-Wilson lattice, we have been led to second order $q$-difference operators $W$ that we have called rational Heun operators. Their relation with the one-particle restriction of the Ruijsenaars-van Diejen Hamiltonians denoted  $A^{(1)}$ by Takemura was established. This supplements the Heun operator of Askey-Wilson (AW) type obtained in \cite{BTVZ}  by considering $q$-difference operators with a raising action on \textit{polynomials} defined on the Askey-Wilson lattice. This last operator in \cite{BTVZ} was shown here to result as a limit of $A^{(1)}$. For the Heun operator of AW type, one had the benefit of the bispectral context provided by the Askey-Wilson polynomials. This allowed for an alternative derivation of the $q$-Heun operators of AW type through tridiagonalization and to cast this $q$-Heun operators within a significant cubic algebra. It would certainly be of relevance to identify algebraic structures that would encompass the rational Heun operators introduced here. Regarding special functions, we also stressed in our analysis, the occurrence of Wilson biorthogonal rational functions as solutions of generalized eigenvalue problems defined in terms of rational Heun operators.

The Heun operators attached to the big and little $q$-Jacobi polynomials that live on $q$-linear lattices have already been constructed in \cite{BVZ_Heun} using both the raising and tridiagonalization approaches. They coincide with Takemura's degenerations $A^{(3)}$ and $A^{(4)}$ of the one-particle Ruijsenaars-van Diejen Hamiltonian. Special cases of the Pastro polynomials, which are examples of biorthogonal polynomials of Laurent type, were seen to occur in this context again as solutions of associated GEVP. We are reserving to an upcoming publication the study of the rational Heun operators corresponding to functions whose poles are distributed on the $q$-linear lattice and the determination of the special functions that solve the corresponding GEVP. It would be enlightening to elucidate the limit relations on the one hand, between the $Ws$ of this paper and these rational Heun operators defined for poles on $q$-linear grids and on the other hand, between the Wilson rational functions and those that will occur in the $q$-linear case. We may suspect that a connection with the operator $A^{(2)}$ of Takemura will also emerge.

Finally, it would be of great interest to understand in the future how all the considerations developed in the present paper could extend in the exploration of Heun operators in many variables.

\setcounter{equation}{0}

\bigskip\bigskip
{\Large\bf Acknowledgments}

The authors are indebted to V.~Spiridonov for drawing their attention to the reference \cite{Spiridonov}. 
The work of S.T. is partially supported by JSPS KAKENHI Grant Numbers 19H01792, 17K18725. 
The research of L.V. is funded in part by a discovery grant from the Natural Sciences and Engineering Research
Council (NSERC) of Canada. The work of A.Z. is supported by the National
Science Foundation of China (Grant No.11771015).

\vspace{15mm}

\bb{99}

\bi{Atk} F.~V.~Atkinson, {\it Multiparameter Eigenvalue Problems}, Academic Press, New York, London, 1972.

\bi{BP} P.~Baseilhac and R.~A.~Pimenta, {\it Diagonalization of the Heun-Askey-Wilson operator, Leonard pairs and the algebraic Bethe ansatz}, arXiv: 1909.02464

\bi{BVZ_Heun}  P.~Baseilhac, L.~Vinet and A.~Zhedanov, {\it The $q$-Heun operator of big $q$-Jacobi type and the $q$-Heun algebra}, Ramanujan J. (2019), doi:10.1007/s11139-018-0106-8, arXiv:1808.06695.

\bi{BTVZ} P.~Baseilhac, S.~Tsujimoto, L.~Vinet and A.~Zhedanov, {\it The Heun-Askey-Wilson algebra and the Heun operator of Askey-Wilson type}, 
Ann. Henri Poincar\'e {\bf 20} (2019), 3091--3112, arXiv:1811.11407.

\bi{CVZ} N.~Cramp\'e, L.~Vinet and A.~Zhedanov, {\it Heun algebras of Lie type}, Proc.~Amer.~Math.~Soc. (2019), doi:10.1090/proc/14788, arXiv: 1904.10643.

\bi{Diejen} J.~F. van Diejen , {\it Integrability of difference Calogero Moser systems}, J. Math. Phys. {\bf 35} (1994), 2983--3004.

\bi{GM} D.~Gupta and D.~Masson, {\it Contiguous Relations, Continued Fractions and Orthogonality}, 
Trans. Amer. Math. Soc. {\bf 350} (1998), 769--808, arXiv:math/9511218v1

\bi{GVZ_Heun} F.~A.~Gr\"unbaum, L.~Vinet and A.~Zhedanov, {\it Tridiagonalization and the Heun equation}, J. Math. Physics {\bf 58}, 031703 (2017), arXiv:1602.04840.

\bi{GVZ_band} F.~A.~Gr\"unbaum, L.~Vinet and A.~Zhedanov, {\it Algebraic Heun operator and band-time limiting}, Commun. Math. Phys. {\bf 364} (2018), 1041--1068, arXiv:1711.07862.

\bi{Hahn_Heun} W.~Hahn, {\it On linear geometric difference equations with accessory parameters}, Funkcial.
Ekvac. {\bf 14} (1971), 73--78.

\bi{KH} Y. Komori and K. Hikami, {\it Quantum integrability of the generalized elliptic Ruijsenaars models}, J. Phys. A Math. Gen. 30 (1997), 4341--4364.

\bi{K} G. Kristensson, {\sl Second Order Differential Equations}, (2010), Springer, New York, NY.

\bi{NRY} M. Noumi, S. Ruijsenaara and Y. Yamada,  {\it The elliptic Painlev\' e Lax equation vs. van Diejen's 8-coupling elliptic Hamiltonian}, arXiv:1903.09738

\bi{Ronveaux} A. Ronveaux (Ed.), {\it Heun's Differential Equations}, Oxford University Press, Oxford, 1995.

\bi{Ruijsenaars} S.~N.~M. Ruijsenaars, {\it Integrable $BC_N$ analytic difference operators: hidden parameter symmetries and eigenfunctions},
in New Trends in Integrability and Partial Solvability, NATO Sci. Ser. II Math. Phys. Chem.,
Vol.132, Kluwer, Dordrecht, 2004, 217--261.

\bi{Sleeman} B.~D.~Sleeman, {\it Multiparameter Spectral Theory in Hilbert Space}, J.Math.Anal.Appl., {\bf 65} (1978), 511--530.

\bi{Spiridonov} V.~P.~Spiridonov, 
{\it Elliptic hypergeometric functions and Calogero-Sutherland-type models}, 
Theor. Math. Phys. {\bf 150} (2007), 266--27.

\bi{Takemura_deg} K.~Takemura, 
{\it Degenerations of Ruijsenaars-van Diejen operator and  $q$-Painleve equations},
J. Integrable Syst. {\bf 2} (2017) , xyx008, 27 pages, arXiv:1608.07265.

\bi{Takemura} K.~Takemura {\it On $q$-deformations of Heun equation}, 
SIGMA {\bf 14} (2018), 061, 16 pages, arXiv:1712.09564.

\bi{Tur_quasi} A. Turbiner, {\it One-Dimensional Quasi-Exactly Solvable Schr\"odinger Equations}, Physics Reports {\bf 642} (2016) 1--71.  arXiv:1603.02992.

\bi{VZ_HH} L.~Vinet and A.~Zhedanov, {\it The Heun operator of Hahn type}, Proc. Amer. Math. Soc. {\bf 147} (2019), 2987--2998, arXiv:1808.00153.

\end{thebibliography}%

\end{document}